\documentclass[12pt]{iopart}

\usepackage{iopams}

\usepackage{graphicx}
\usepackage{dcolumn}
\usepackage{float}
\usepackage{iopams}

\begin{document}

\title{\boldmath Simulation of a hump structure in the optical scattering rate within a generalized Allen formalism and its application to copper oxide systems \unboldmath}

\author{Jungseek Hwang}
\address{Department of Physics, Sungkyunkwan University, Suwon, Gyeonggi-do 440-746, Republic of Korea}

\ead{jungseek@skku.edu}

\date{\today}

\begin{abstract}
We propose a possible way to simulate a hump structure in the optical scattering rate. The optical scattering rate of correlated charge carriers can be defined within an extended Drude model formalism. When some electron and hole doped copper oxide systems are in the spin density or charge density wave phases they show hump structures in their optical scattering rates. The hump structures have not yet been simulated and understood clearly. We are able to simulate the hump structure by using a peak followed by a dip feature in the normalized density of states within a generalized Allen formalism. We observe that reversing the order of the dip and peak gives completely different features in the optical scattering rate; a peak-dip (dip peak) results in a hump (a valley) in the scattering rate. We also obtain the real part of the optical conductivity and reflectance spectra from the simulated optical scattering rate and compare them with published experimental spectra. From these comparisons we conclude that the peak-dip order can give the hump structure, which is observed experimentally in copper oxide systems. Finally we fit two published optical spectra with our new model and discuss our results and the possible origin of the dip or peak features in the normalized density of states.
\end{abstract}

\maketitle

\section{Introduction}

High temperature cuprates were discovered in 1986\cite{bednorz:1986}. Since that time, many spectroscopic studies with different experimental techniques have been done on this intriguing material system\cite{carbotte:2011}. However, the microscopic mechanism of the superconducting phenomenon of the material has not yet been figured out. The electron doped cuprate system has been studied relatively less than the hole doped, because of its lower transition temperature, even though its superconducting mechanism is believed to be the same as for the hole doped. Uderdoped electron cuprates show a spin density wave (SDW) in mid infrared range, which appears as a partial suppression in the density of states\cite{parkSR:2007,yasuoka:1989}. This SDW seems to be correlated to a hump structure in the optical scattering rate\cite{onose:2004,wang:2006}. This hump feature has not yet been simulated and clearly understood. Similar hump features have also been observed in the optical scattering rate of underdoped hole cuprate systems when they contain the charge density wave (CDW) or charge stripes\cite{dumm:2002,homes:2012} even though this have ever been pointed out explicitly so far as the author knows.

Here we propose a possible way to simulate the hump structure within a generalized Allen formalism or the Sharapov-Carbotte formula\cite{allen:1971,mitrovic:1985,sharapov:2005}, which can be applied to material systems both at finite temperature and with non-constant density of states. We perform model calculations to get the optical scattering rates by using the generalized Allen formula with four differently shaped densities of states (peak only, dip only, peak-dip, and dip-peak cases) and two different electron-boson spectral functions (the Millis, Monien, and Pines (MMP) type antiferromagnetic spin fluctuation model\cite{millis:1990} and the MMP + a sharp Gaussian mode). From these calculations we find that to get the hump structure in the scattering rate we must have both peak and dip structures in the normalized density of states (DOS) and the peak and dip should be located at a finite frequency. We also observe that the different orders of the peak and dip result in different features in the optical scattering rate; while a peak-dip order gives a hump structure, as observed in some experiment, a dip-peak order gives a valley (or a suppression, which is similar to the usual pseudogap\cite{hwang:2012} model except now the dip starts from a finite frequency. We emphasize that the peak-dip order is a new model and different from the pseudogap model which has been used to extract the electron-boson spectral function from the optical scattering rate of underdoped cuprates\cite{hwang:2008,hwang:2006}. While the pseudogap model gives a suppression, this new peak-dip model gives an enhancement (a hump) in the scattering rate. We also obtained the optical conductivity and reflectance spectra from the calculated optical scattering rates and compare the spectra with experimental data. From these comparisons we conclude that the peak-dip model can be applied to optical data with the hump structure. We apply this model to extract the electron boson spectral functions from the measured spectra of two copper oxide systems and discuss the results obtained.

In section 2 we introduce the general formalism used in this study. In section 3 we show our simulated results obtained using the formalism with various model density of states. In section 4 we calculate the optical conductivity and reflectance from the calculated optical scattering rate and compare the results with published experimental spectra. In section 5 we apply our new model to the measured spectra of two copper oxides. In section 6 we discuss our results and possible origin of the dip and peak structures in the density of states.

\section{General formalisms}

Optical spectroscopy has been used to study strongly correlated electron systems including high temperature superconducting materials\cite{basov:2005}. Typically we measure the reflectance spectra, $R(\omega)$ of superconducting or metallic samples at various temperatures above and below some characteristic temperature including the superconducting transition temperature. In general, the reflectance spectrum is quite a complicated physical quantity from the theoretical or analytical points of view. Therefore we need to extract simpler quantities from the measured reflectance spectrum. These simpler quantities are known as the optical constants, like the dielectric constant, the index of refraction, and so on. To get optical constants from the measured reflectance spectrum, we usually rely on a Kramers-Kronig (K-K) integral relation\cite{wooten}. To perform the K-K integral we have to extrapolate the measured reflectance spectrum (which is known only in a finite spectral range because of experimental limitations) to both zero and infinite frequency guided by appropriate physical models since the integral range in the K-K relation is from 0 to $\infty$. Once we get the corresponding phase of $R(\omega)$ through the K-K analysis we can calculate other optical constants using well-known relationships between them\cite{wooten}.

For further analysis we rely on an extended Drude model formalism\cite{allen:1977,puchkov:1996,timusk:1999,hwang:2004}. The extended Drude conductivity can be written as follows:
\begin{equation}\label{eq1}
\tilde{\sigma}(\omega) = i\frac{\omega_p^2}{4\pi}\frac{1}{\omega+[-2\tilde{\Sigma}^{op}(\omega)+i1/\tau_{imp}]}
\end{equation}
where $\omega_p^2$ is a square of the plasma frequency, which is proportional to the number density of charge carriers in a conducting material, more explicitly $\omega_p^2 \equiv 4\pi n_{e} e^2/m^{*}$, where $n_e$ is the number density of free charge carriers, $e$ the unit charge and $m^*$ the effective mass of the charge carrier, also known as the band mass. $1/\tau_{imp}$ is the elastic scattering rate of charge carriers due to scattering off impurities. Here $\tilde{\Sigma}^{op}(\omega)$ is the optical self energy, which is a complex function. This quantity carries important information on correlations between the charge carriers. The correlation affects the optical processes: transitions from filled to empty states. The optical self energy can be related to the mass enhancement factor ($\lambda^{op}(\omega)$) and optical scattering rate ($1/\tau^{op}(\omega)$), $\tilde{\Sigma}^{op}(\omega)=\Sigma_1^{op}(\omega)+i\Sigma_2^{op}(\omega) = -[\omega\lambda^{op}(\omega)+i1/\tau^{op}(\omega)]/2$. The real and imaginary parts of the optical self energy are also related through a Kramers-Knonig transformation, {\it i.e.} the complex optical self energy is a causal function.

Since the optical self energy carries information on the interaction between charge carriers, it can be related to the electron-boson spectral function\cite{allen:1971}, $\alpha^2F(\omega)$, which describes electron-electron interactions by exchanging a mediating boson. The electron-boson spectral function is also called the Eliashberg function. The most generalized relationship between the optical self energy and the electron-boson spectral function within Allen's formalism was derived by Sharapov and Carbotte\cite{sharapov:2005} and can be written as follows:
\begin{eqnarray}\label{eq2}
-2\Sigma_2^{op}(\omega,T)&\equiv&\frac{1}{\tau^{op}(\omega,T)} \\\nonumber
&=&\!\!\frac{\pi}{\omega}\int_{0}^{\infty}\!\!\!\!\!d\Omega \alpha^2F(\Omega)\int_{-\infty}^{\infty}\!\!\!\!\!dz[N(z-\Omega)\!\!+\!\!N(-z+\Omega)] \\\nonumber
&\times& [n_B(\Omega)+1-f(z-\Omega)][f(z-\omega)-f(z+\omega)]
\end{eqnarray}
where $N(z)$ is the normalized density of states and $f(\omega)$ and $n_{B}(\omega)$ are the Fermi-Dirac and Bose-Einstein distribution functions respectively. We note that we use a symmetrized DOS as an approximation. Here we neglect vertex corrections and we also assume that $\alpha^2F(\Omega)$ is independent of momentum. If retarded interactions among electrons contribute to the formation of the cooper pairs in exotic superconductors including high temperature superconductors, the electron-boson function may play the role of glue for the pairing. But there is also experimental evidence which shows that nonretarded interactions can also contribute to the pairing in cuprates\cite{Mansart:2013}. This generalized Allen model can be applied to analyze correlated electron systems with non-constant density of states like a pseudogap, which is a partial gap near the Fermi energy\cite{timusk:1999}. In cuprates the $\alpha^2F(\Omega)$ is related to the antiferromagnetic spin fluctuations. The non-constant density of states features (peak, dip, or a combination of the two) in DOS and the spin fluctuations are interrelated in a feedback process and appear in the optical scattering rate as in Eqn. (\ref{eq2}). We can obtain the electron-boson spectral function from the measured optical scattering rate by solving this integral equation numerically. In this numerical process one has to know the exact shape of the pseudogap since the pseudogap affects the resulting electron-boson spectral function significantly\cite{hwang:2006,hwang:2012}. This numerical process is known as an inversion process. There are two different numerical approaches to solve the integral equation: one is a least square fitting method\cite{dordevic:2005,hwang:2006} and the other is a maximum entropy method\cite{schachinger:2006}.

\section{Simulation of hump structure in the optical scattering rate}

\begin{figure}[t]
  \vspace*{-1.0 cm}%
  \centerline{\includegraphics[width=4.0 in]{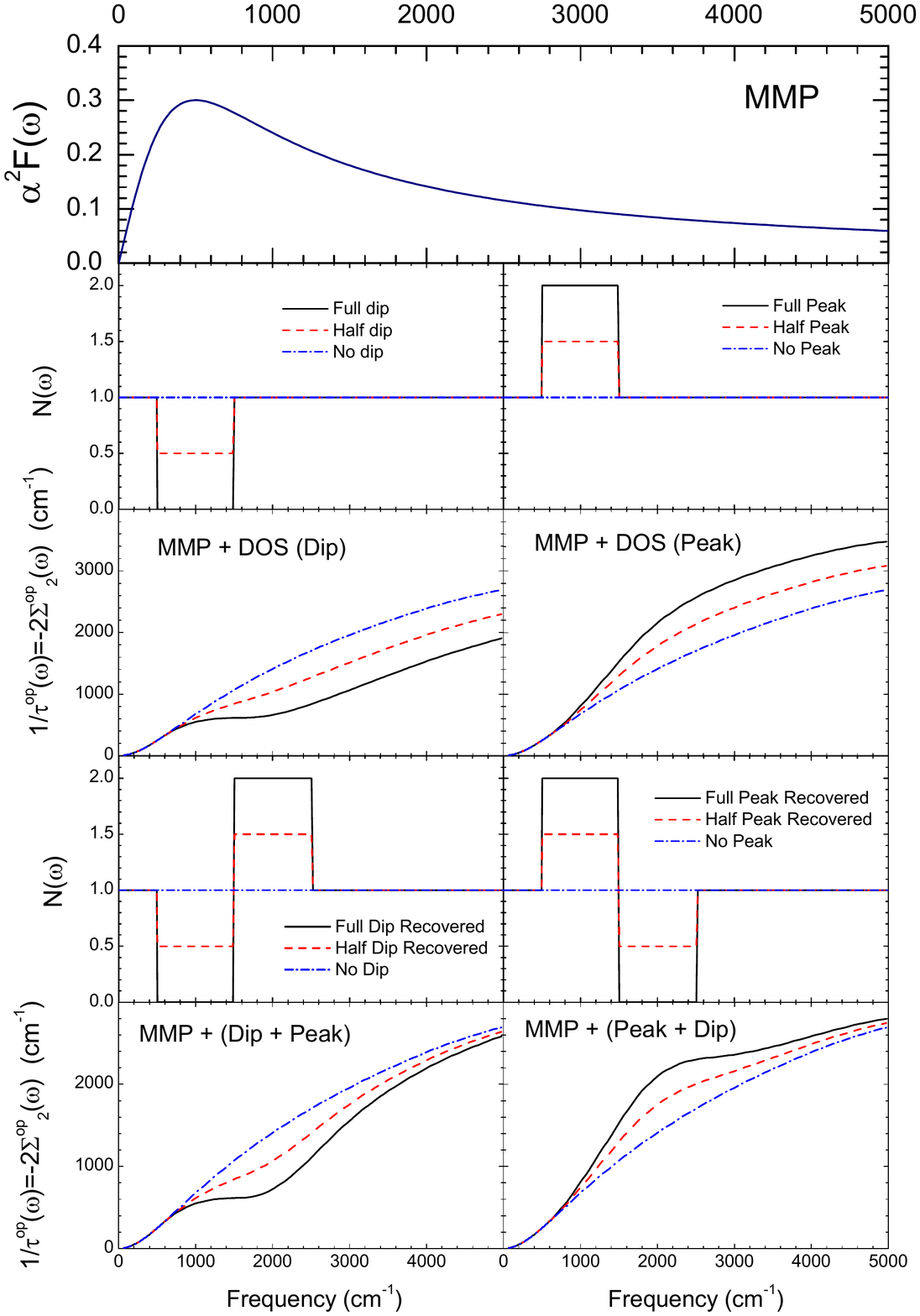}}%
  \vspace*{-0.9 cm}%
\caption{(Color online) The electron-boson spectral function, $\alpha^2F(\omega)$, consists of the MMP type spin fluctuation mode only (top frame). The input parameter functions for the electron-boson spectral function (top frame) and for two different normalized density of states ($N(z)$) cases (the 2nd row) and the resulting optical scattering rate (the 3rd row). The input parameter functions for the electron-boson spectral function (top frame) and for two different $N(z)$ cases (the 4th row) and the resulting optical scattering rate (the last row). In this case the DOS loss (gain) in the dip (peak) is recovered completely.}
 \label{fig1}
\end{figure}

\begin{figure}[t]
  \vspace*{-1.0 cm}%
  \centerline{\includegraphics[width=4.0 in]{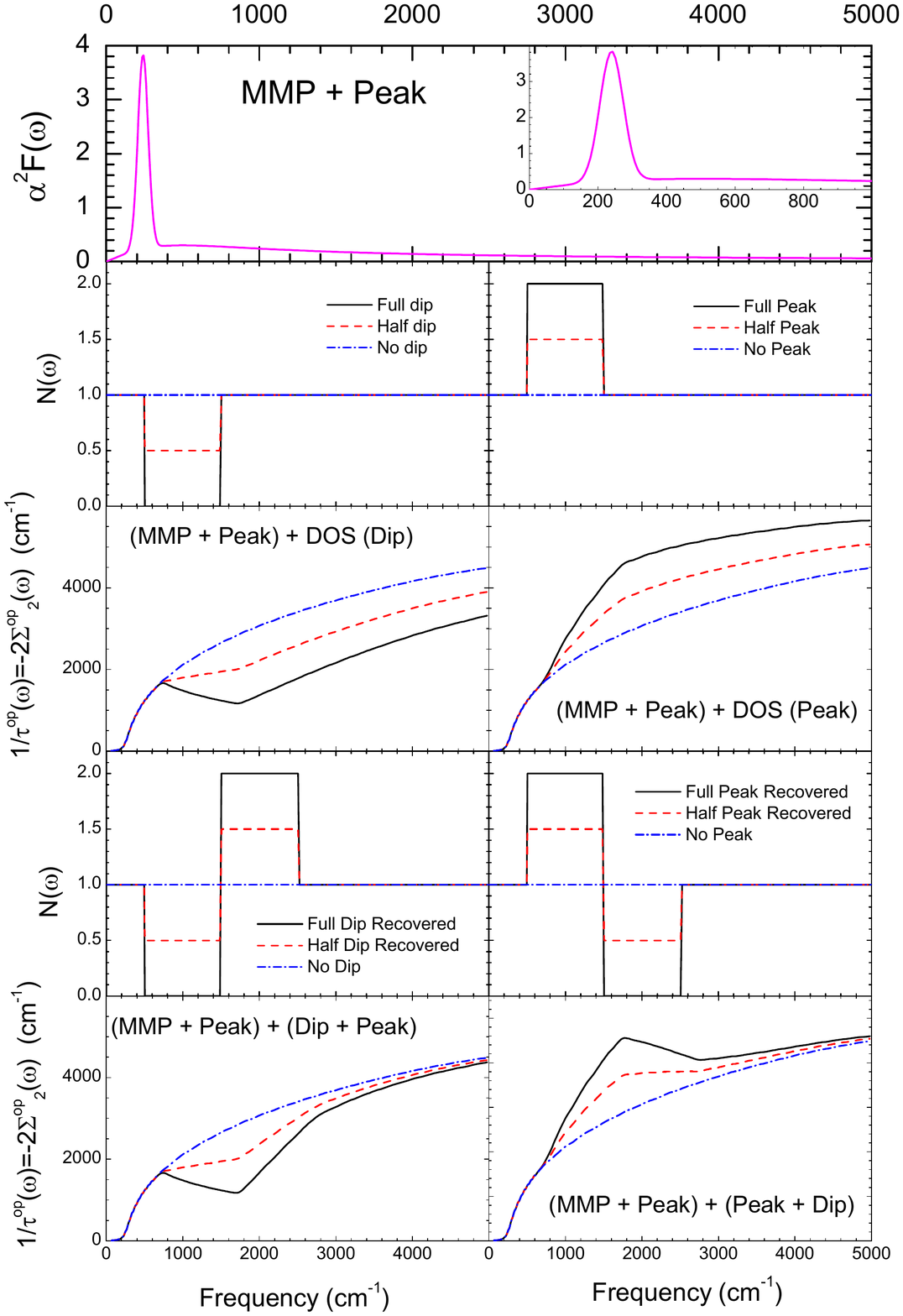}}%
  \vspace*{-0.9 cm}%
\caption{(Color online) The electron-boson spectral function, $\alpha^2F(\omega)$, consists of the MMP type spin fluctuation mode and a sharp Gaussian peak (top frame and also see in the text). The input parameter functions for the electron-boson spectral function (top frame) and for two different normalized density of states ($N(z)$) cases (the 2nd row) and the resulting optical scattering rate (the 3rd row). In the inset of the middle frame we show $\alpha^2F(\omega)$ in low frequency region. The input parameter functions for the electron-boson spectral function (top frame) and for two different normalized density of states ($N(z)$) cases (the 4th row) and the resulting optical scattering rate (the last row). $\alpha^2F(\omega)$ consists of  the MMP type spin fluctuation mode and a sharp Gaussian peak. In the inset of the middle frame we show $\alpha^2F(\omega)$ in low frequency region. The density of states loss (gain) in the dip (peak) is recovered completely.}
 \label{fig2}
\end{figure}

Many inversion studies of the optical scattering rates of copper oxide superconductors have been done\cite{hwang:2006,schachinger:2006,hwang:2007,yang:2009}. There is a theoretical literature, which shows hump structures in the calculated optical scattering rate\cite{knigavko:2005}. However, the hump structures appear in the scattering rate because of finite band effects. Here we show that hump structures can also be realized simply by manipulating the shape of the non-constant density of states within the generalized Allen formalism. From the integral equation, Eqn. (\ref{eq3}), we see that the scattering rate is an increasing function of frequency if the density of states remains larger or equal to the normalized value, 1.0 at any energy. It can decrease only when the density of states becomes less than 1.0. We performed our model calculations at 0 K for simplicity. At finite temperatures the qualitative properties will remain the same as those at 0 K. We used two different electron-boson spectral function ($\alpha^2F(\omega)$) models: the MMP model\cite{millis:1990} and the MMP + a sharp Gaussian model. These models are well-known typical shapes of the electron-boson spectral function in cuprate systems\cite{schachinger:2000,hwang:2006}.

In Fig. \ref{fig1} we show the optical scattering rates obtained using Eqn. (\ref{eq3}) with the input electron-boson spectral function $\alpha^2F(\omega)$ (top frame) and the normalized density of states $N(\omega)$ (the 2nd and 4th rows). We use the MMP antiferromagnetic spin fluctuation model for the electron-boson function, {\it i.e.} $\alpha^2F(\omega)=\frac{A_s \omega}{\omega^2+\omega_{sf}^2}$ where $A_s =$ 300 cm$^{-1}$ and $\omega_{sf} =$ 500 cm$^{-1}$ are the amplitude and the characteristic frequency of the spin fluctuation MMP mode, respectively. $N(\omega)$ has a rectangular dip (in left column of the 2nd row) or a rectangular peak (in right column of the 2nd row) with the same width ($W$) 1000 cm$^{-1}$ in a spectral range between $\omega_{Low}$ (= 500 cm$^{-1}$) and $\omega_{High}$ (= 1500 cm$^{-1}$), where $\omega_{Low}$ and $\omega_{High}$ are, respectively, a lower frequency edge (or an onset frequency) and a higher frequency one of the rectangular dip (or peak). We calculate the optical scattering rates for three different depths (heights) of the dip (peak): the black solid curve is for a full dip/peak, the red dashed line for a half dip/peak, and the blue dot-dashed curve for no dip/peak ({\it i.e.} a constant density of states). The dip (peak) in the normalized DOS introduces a abrupt suppression (a sharp increase) in the optical scattering rate with its onset frequency near 1000 cm$^{-1}$ ($\simeq \omega_{sf}+\omega_{Low}$) because of reduced (increased) amount of DOS involved in the scattering process. We note that the onset frequency of the suppression or sharp increase is shifted by roughly the characteristic frequency ($\omega_{sf}$) of the MMP mode from the onset frequency ($\omega_{Low}$) of the dip or peak. In the 4th and last rows the DOS loss (gain) in the dip (peak) is recovered completely by a peak (dip) right above the dip (peak) as shown in the the 4th row. The recovery in DOS ensures that the scattering rate at high frequencies is recovered. We observe that different orders of the peak and the dip give different shapes for the resulting scattering rates; while the dip-peak order (in left column) gives a well-defined valley the peak-dip order (in right column) gives a well-defined hump in the scattering rate as compared with the flat DOS case, displayed as the blue dash-dotted curve. The minimum of the valley (or the maximum of the hump) is located near 2000 cm$^{-1}$ which is the sum of the three characteristic energy scales, {\it i.e.} $\simeq\omega_{sf}+\omega_{Low}+W$.

In Fig. \ref{fig2} we display similar results as in Fig. \ref{fig1} with a different electron-boson spectral function, $\alpha^2F(\omega)$. Here we have both the MMP mode and a sharp Gaussian peak (top frame) $\alpha^2F(\omega)=\frac{A_s \omega}{\omega^2+\omega_{sf}^2}+\frac{A}{\sqrt{2 \pi}(d/2.35)}e^{-(\omega-\omega_0)^2/[2(d/2.35)^2]}$ where $A_s =$ 300 cm$^{-1}$, $\omega_{sf}=$ 500 cm$^{-1}$ are the amplitude and the characteristic frequency of the spin fluctuation MMP mode, respectively. $A =$ 300 cm$^{-1}$, $\omega_0 =$ 240 cm$^{-1}$, and $d =$ 80 cm$^{-1}$ are the amplitude, the center frequency, and the width of the Gaussian peak, respectively. The Gaussian peak can be modeled as the well-known magnetic resonance mode which was observed first in inelastic neutron scattering experiments\cite{rossat:1991,fong:2000,stock:2004}. Since the magnetic resonance has strong momentum dependence current vertex correction may need to be included in the model. We ignore the vertex correction since it is none trivial to include these in our model but this is not a major issue for this paper. In the 2nd and 3rd rows, we also observe that the rectangular dip (peak) in the normalized density of states introduces a abrupt suppression (a sharp increase) in the optical scattering rate. The sharp Gaussian peak enhances these features in the scattering rate. The sharp Gaussian peak introduces kinks ({\it i.e.} sudden slope changes) in the scattering rate instead of the smooth slope changes which we observed previously for the MMP only case. The kinks indicate the characteristic energy scales of the input parameter functions ($N(\omega)$ and $\alpha^2F(\omega)$) clearly. The onset frequency of the abrupt suppression (or the sharp increase) in the scattering rate appears near 740 cm$^{-1}$ which is the sum of two characteristic energy scales: the onset frequency ({\it i.e.} $\omega_{Low} =$ 500 cm$^{-1}$) of the rectangular dip (or peak) in the DOS and the Gaussian peak frequency ({\it i.e.} $\omega_0 =$ 240 cm$^{-1}$) in $\alpha^2F(\omega)$. In the 4th and last rows, DOS loss (gain) in the rectangular dip (peak) is recovered by a peak (dip) right above the dip (peak) as shown in the the 4th row. Because of the recovery in the density of states the scattering rates come together at high frequencies. We also see a valley (a hump) in the scattering rate for the dip-peak (peak-dip) case; as we have pointed out already we have a kink inside the valley or on top of the hump instead of a smooth slope change because of the sharp Gaussian feature in $\alpha^2F(\omega)$. The frequency of the kink on top of the hump (or bottom of the valley) in the scattering rate is roughly 1740 cm$^{-1}$, which is the sum of three characteristic energy scales: the onset frequency of the rectangular dip (peak) ({\it i.e.} $\omega_{Low} =$ 500 cm$^{-1}$) in the DOS, the width ({\it i.e.} $W =$ 1000 cm$^{-1}$) of the dip or peak and the Gaussian peak frequency ({\it i.e.} $\omega_0 =$ 240 cm$^{-1}$) in $\alpha^2F(\omega)$. We see the characteristic energy scales clearly in the simulated scattering rates. It may be possible to get these characteristic energy scales ($\omega_{Low}$, $\omega_{High}$, $W$, $\omega_{sf}$, and $\omega_0$) from the measured scattering rate, which would provide useful information when one analyzes measured data using our model approach.

\begin{figure}[t]
  \vspace*{-1.0 cm}%
  \centerline{\includegraphics[width=4.0 in]{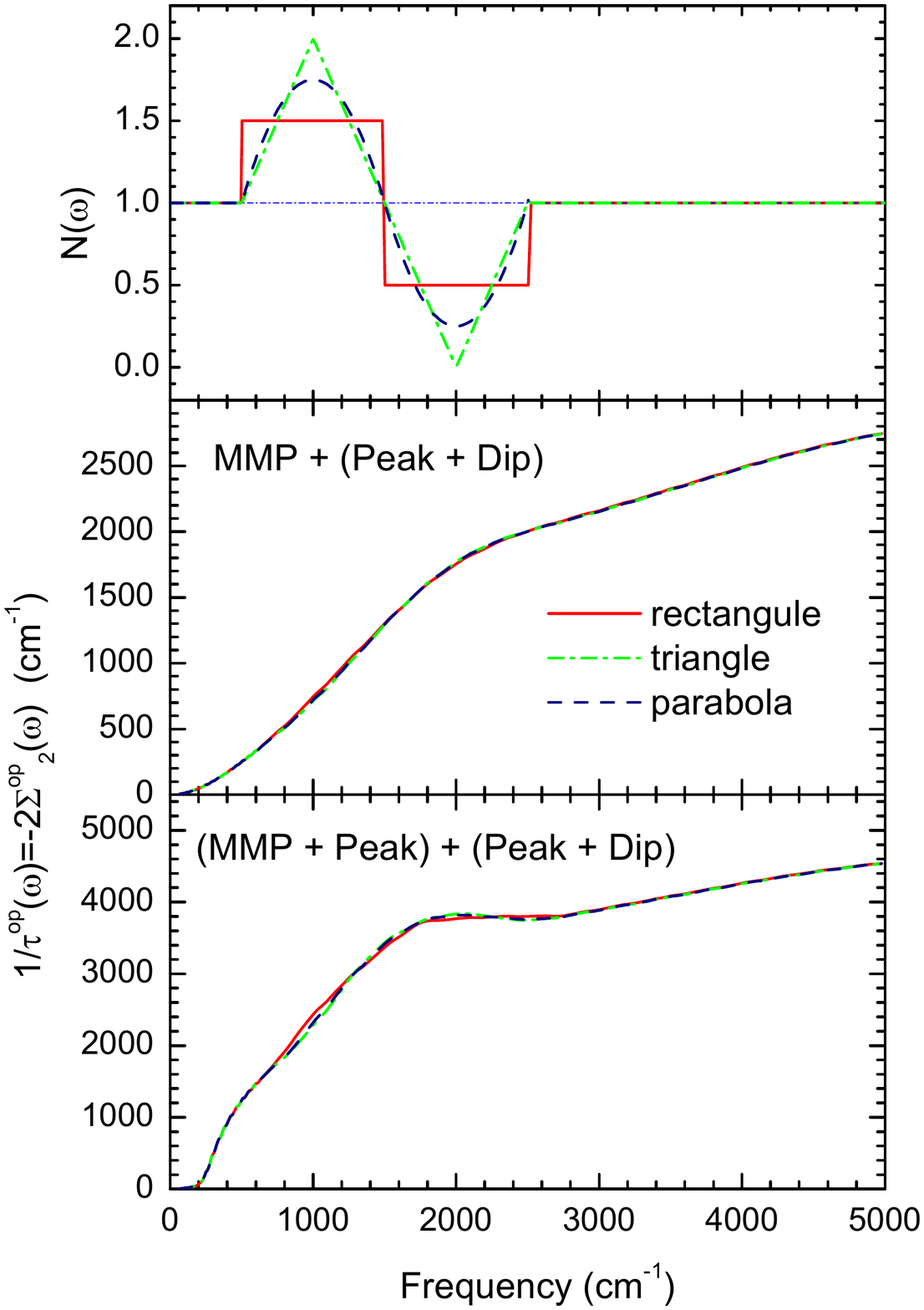}}%
  \vspace*{-0.9 cm}%
\caption{(Color online) A shape sensitivity of the peak-dip feature in DOS to the resulting optical scattering rate. In the top frame three differently shaped peak-dip features are shown. In the middle frame we display corresponding three scattering rates for the MMP electron-boson function case. In the bottom frame we show corresponding three scattering rates for the MMP plus the sharp Gaussian electron-boson function case.}
 \label{fig2a}
\end{figure}

We also checked the shape sensitivity of the peak-dip feature in the resulting optical scattering rate. We simulate the optical scattering rate with three different shapes of the peak-dip structure (rectangle, triangle, and parabola) as shown in the top frame of Fig \ref{fig2a}. The resulting scattering rates are shown in middle and bottom frame for the MMP bosonic mode and the MMP plus the sharp Gaussian mode cases, respectively. We can see in the resulting scattering rates that the shape dependence is negligible.

\section{Optical conductivity and reflectance spectra calculated from the simulated scattering rates}

\begin{figure}[t]
  \vspace*{-1.0 cm}%
  \centerline{\includegraphics[width=4.0 in]{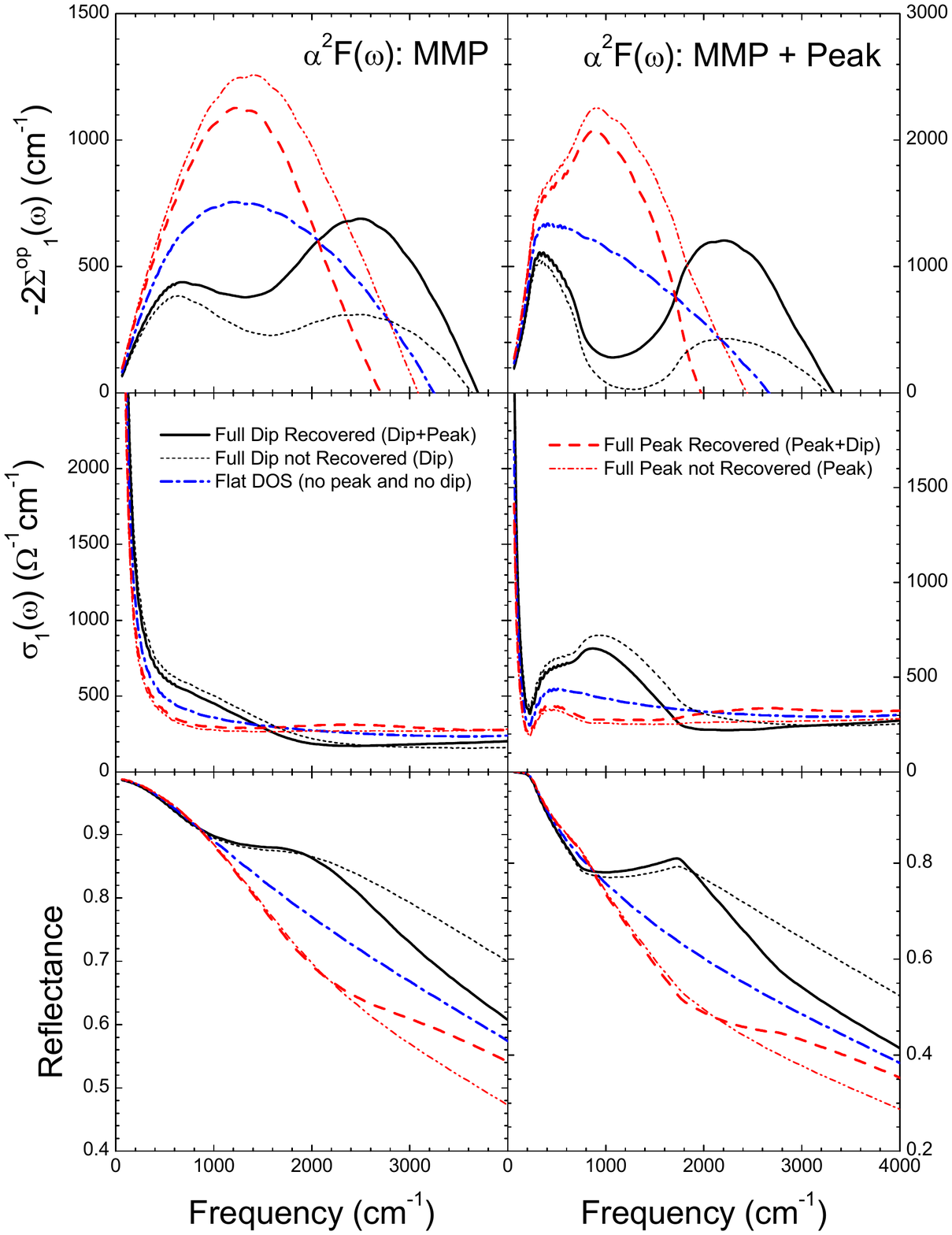}}%
  \vspace*{-0.9 cm}%
\caption{(Color online) (Left column) The calculated real part of the optical self energy (top frame), the real part of the optical conductivity (middle frame), and reflectance spectra (bottom frame) for the MMP electron-boson mode and five different representative normalized DOS cases. (Right column) The calculated spectra for the MMP plus the sharp Gaussian electron-boson mode and five different normalized DOS cases.}
 \label{fig3}
\end{figure}

\begin{figure}[t]
  \vspace*{-1.0 cm}%
  \centerline{\includegraphics[width=4.0 in]{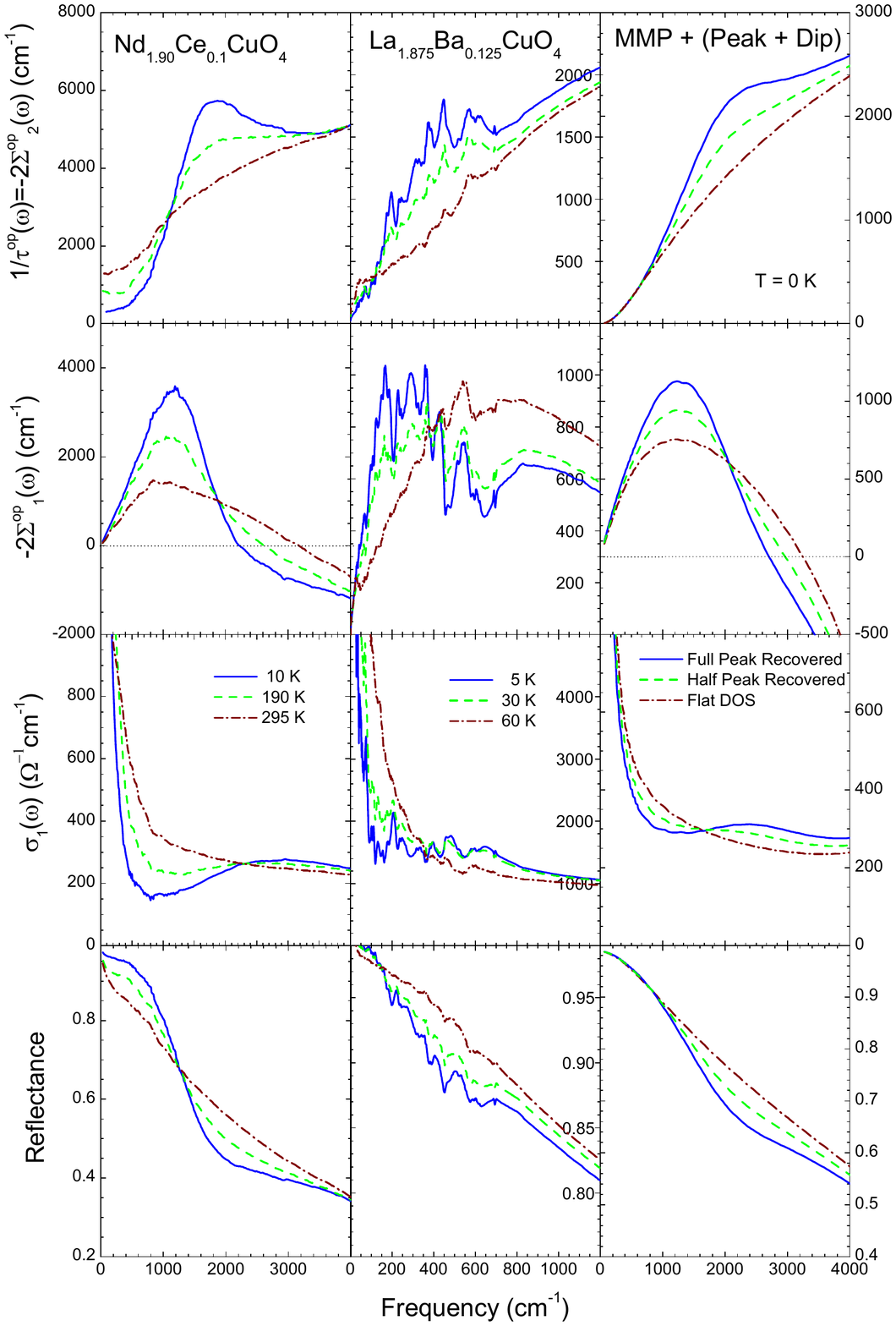}}%
  \vspace*{-0.9 cm}%
\caption{(Color online) Comparison of the simulated results (right column) with measured electron underdoped data of Nd$_{1.90}$Ce$_{0.10}$CuO$_4$ (left column)\cite{wang:2006} and with measured hole underdoped data of La$_{1.875}$Ba$_{0.125}$CuO$_4$ (middle column)\cite{homes:2006,homes:2012}. In the middle column the scales on the vertical axis are shown on the right side.}
 \label{fig4}
\end{figure}

We obtain the corresponding optical conductivity and reflectance spectra from the simulated optical scattering rates of the previous section and compare them with published data. To get the reflectance from the calculated optical scattering rate we have to go through a series of mathematical procedures. Once we know the optical scattering rate ({\it i.e.} the imaginary part of the optical self energy) we can calculate the real part of the optical self energy using a Kramers-Kronig (K-K) transformation. Then we can calculate the optical conductivity using the extended Drude model, {\it i.e.} Eqn. (\ref{eq2}). More explicitly we can write down the real and imaginary parts of the optical conductivity as a function of the real and imaginary parts of the optical self energy as follows:
\begin{eqnarray}\label{eq3}
\sigma_1(\omega)\!\!&=&\!\!\frac{\omega_p^2}{4 \pi}\frac{-2\Sigma_2^{op}(\omega)+1/\tau_{imp}}{[-2\Sigma_1^{op}(\omega)+\omega]^2+[-2\Sigma_2^{op}(\omega)+1/\tau_{imp}]^2} \\\nonumber
\sigma_2(\omega)\!\!&=&\!\!\frac{\omega_p^2}{4 \pi}\frac{-2\Sigma_1^{op}(\omega)+\omega}{[-2\Sigma_1^{op}(\omega)+\omega]^2+[-2\Sigma_2^{op}(\omega)+1/\tau_{imp}]^2}
\end{eqnarray}
We can easily obtain the reflectance spectrum from the complex optical conductivity\cite{hwang:2007a}. For these calculations we used $1/\tau_{imp} =$ 100 cm$^{-1}$, $\omega_p =$ 10,000 cm$^{-1}$, and $\epsilon_{\infty} =$ 1.0. In practice we need to use an appropriate value for $\epsilon_{\infty}$ since we have to include contributions from high energy absorption features such as the ionic background, the interband absorptions and so on\cite{hwang:2007a}. In that case $\epsilon_{\infty}$ is always larger than 1.0.

We display three representative calculated quantities: the real part of the optical self energy (top frames), the real part of the optical conductivity (middle frames), and reflectance (bottom frames) in Fig \ref{fig3}. Left column is for the MMP model of $\alpha^2F(\omega)$ and five representative different density of states (DOS) taken from the cases we discussed in the previous section: flat DOS, full dip only, full peak only, full dip+full peak (completely recovered), and full peak+full dip (completely recovered). First of all, as we expect, we see significant differences between full dip only (short dashed thin black curve) and full peak only (dot-dot dashed thin red curve) cases and between the dip+peak (solid thick black curve) and peak+dip (dashed thick red curve) ones. In the conductivity (the middle frame of the left column) we see a suppression for the peak only and the peak+dip cases at low frequency region and in reflectance (the bottom frame of the left column). We also observe a suppression for the same cases in a frequency range starting from 1000 cm$^{-1}$. While for the peak only case the reflectance is lower than for the flat DOS case at high frequencies, for the peak+dip case the reflectance recovers only slowly to that of the flat case at high frequencies. This frequency dependent behavior results in a valley in the reflectance of the peak+dip case near 2000 cm$^{-1}$ where the hump is in the optical scattering rate (see the black solid curve in the left column of the last row of Fig. \ref{fig1}). This feature in reflectance spectra of the peak+dip case looks similar that of electron underdoped cuprates and hole doped cuprates \cite{dumm:2002,onose:2004,zimmers:2005,homes:2006,wang:2006,homes:2012} (see also Fig. \ref{fig4} for closer comparisons). In the right column we display the same quantities as in the left column but with a different electron-boson spectral function $\alpha^2F(\omega)$, which consists of both the MMP and the sharp Gaussian modes. The quantities show qualitatively similar properties as those in the left column. However, the sharp Gaussian peak gives better defined features as we saw in the optical scattering rate in the previous section. Since charge carriers are strongly correlated we see clearly a boundary between the coherent and incoherent parts of the conductivity near 240 cm$^{-1}$ where the Gaussian peak exists\cite{hwang:2008a}. We also observe similar characteristic energy scales to those observed in the optical scattering rate in the previous section.

In Fig. \ref{fig4} we compare our resulting spectra (the optical scattering rate, the real part of optical self energy, the optical conductivity, and reflectance ordered from top to bottom frames in the right column) from the simulation with measured optical spectra of an electron doped Nd$_{1.90}$Ce$_{0.10}$CuO$_4$ (left column), which is non-superconducting\cite{wang:2006} and has a spin density wave phase\cite{onose:2004} at low temperatures. The simulated data are obtained with the MMP mode of the electron-boson spectral function and three different peak-dip normalized densities of states (full peak+full dip, half peak+half dip, and flat DOS). We modified the original scattering rates slightly in the low frequency region to remove possibly experimental noises. We note that while the experimental data are at finite temperatures, the simulated ones are at the zero temperature. We can see that the two data sets are qualitatively very similar if we ignore finite temperature effects in experimental data; the hump structure appears at low temperature and disappears gradually as the temperature increases up to room temperature. We compare the same simulated spectra (right column) with the measured spectra (middle column) of an underdoped hole La$_{1.875}$Ba$_{0.125}$CuO$_4$, which has a charge ordered or stripe phase\cite{tranquada:2004,homes:2006} below 60 K. We see that the two data sets show similar qualitative properties; the hump structure in the optical scattering rate grows gradually as we reduce the temperature below the charge ordering transition temperature, 60 K. We emphasize that the peak-dip model agrees with the observed experimental spectra. In the next section we apply the peak-dip model to two experimental measured spectra.

\section{Applications to two copper oxides systems}

\begin{figure}[t]
  \vspace*{-1.0 cm}%
  \centerline{\includegraphics[width=4.0 in]{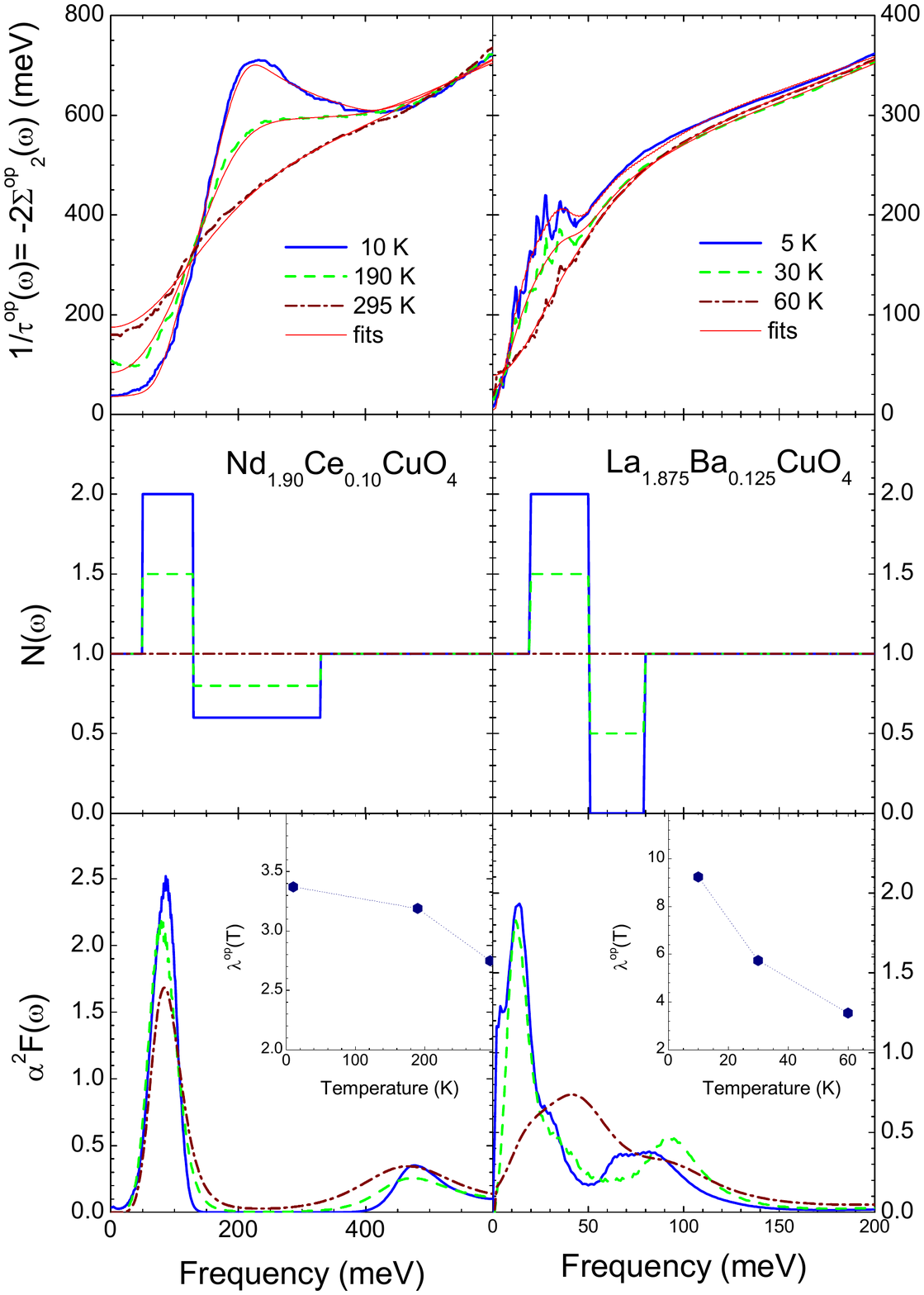}}%
  \vspace*{-0.9 cm}%
\caption{(Color online) (Left column) Application of our model to measured electron underdoped data of Nd$_{1.90}$Ce$_{0.10}$CuO$_4$\cite{wang:2006} and the results. We use a maximum entropy method\cite{hwang:2013} to reconstruct the measured data by using the new model. (Right column) Application of the model to measured hole underdoped data of La$_{1.875}$Ba$_{0.125}$CuO$_4$\cite{homes:2006,homes:2012} and the results.}
 \label{fig5}
\end{figure}

We apply the peak-dip model to analyze in more detail two selected experimental data sets\cite{homes:2006,wang:2006}: Nd$_{1.90}$Ce$_{0.10}$CuO$_4$ (non superconducting) and La$_{1.875}$Ba$_{0.125}$CuO$_4$ ($T_c \simeq$ = 2.4 K). Since both data sets are in the normal state we may use the generalized Allen formula, Eqn. \ref{eq3} to analyze them. In the top frame of the left column of Fig. \ref{fig5} we display the optical scattering rates of the electron doped Nd$_{1.90}$Ce$_{0.10}$CuO$_4$ at three different temperatures (10, 190 and 295 K) and the reconstruction data obtained using a maximum entropy method (MEM)\cite{schachinger:2006,hwang:2012}. The fitting quality is quite good; we are clearly able to simulate the hump structure in the scattering rate. We note that when we fit the spectrum at 10 K we need to add a 35 meV impurity scattering rate to improve the quality of the fit. In the middle frame we show the input normalized densities of states for three different temperatures. The characteristic energies of the peak and dip in the density of states are determined from the characteristic energies and shape of the hump in the optical scattering rate. The input DOS for each temperature is fixed. We note that we need to take a dip which is wider than the peak to fit to the data more tightly. We also note that since the height of the peak and depth of the dip are only roughly guessed in the fittings, the extracted $\alpha^2F(\omega)$'s may not be determined uniquely. In the bottom frame we display the resulting electron-boson spectral function obtained from our MEM analysis. The shape of the electron-boson spectral function looks like that of a typical underdoped hole cuprate\cite{hwang:2011}. The optical mass enhancement factor, $\lambda^{op} \equiv 2\int^{\omega_c}_0 \alpha^2F(\omega)/\omega \:d\omega$, shows a typical temperature dependence as seen in the inset of the bottom frame; as temperature lowers this factor increases\cite{hwang:2011}. Here $\omega_c$ is a cutoff frequency, taken to be 600 meV. Results obtained from the same analysis on the underdoped hole La$_{1.875}$Ba$_{0.125}$CuO$_4$ at three different temperatures (5, 30 and 60 K) are displayed in the right column of Fig. \ref{fig5}. In the top frame we display the optical scattering rates and corresponding fits. We note that the sharp features in the data between 10 and 50 meV are ignored. We use a fixed density of states for each temperature as shown in the middle frame. The resulting electron-boson functions are displayed in the bottom frame. The function shows a strong temperature dependence; as temperature decreases a large amount of spectral weight moves to lower frequency. This behavior causes the mass enhancement factor to increase considerably on lowering the temperature as we can see in the inset of the bottom frame; this is also the typical temperature dependent behavior for the mass factor of hole underdoped cuprates\cite{hwang:2011}. However at our lowest temperature 5 K, the factor is unusually large, {\it i.e.} 9.2. This may be related to the central mode which was observed in a highly underdoped YBa$_2$Cu$_3$O$_{6.353}$ by inelastic neutron scattering experiment\cite{stock:2006}.

We push the analysis a little further with some assumptions. We assume that the extracted $\alpha^2F(\omega)$ contributes to the superconductivity even though the $\alpha^2F(\omega)$ is extracted from the optical scattering rate in the normal state. We also assume that we can use the extended McMillan formula\cite{mcmillan:1968,williams:1989,hwang:2008c} to estimate fictitious superconducting transition temperatures. The McMillan formula can be written as $k_BT_c \cong 1.13 \hbar \omega_{ln} \mbox{exp}[-(1+\lambda^{op})/(g\lambda^{op})]$, where $\omega_{ln}\equiv \mbox{exp}[2/\lambda \int_0^{\omega_c}\mbox{ln}\omega\: \alpha^2F(\omega)/\omega\: d\omega]$ is the logarithmically averaged boson frequency, and $g$ is an adjustable parameter ($g\in[0,1]$), which may allow one to take the d-wave nature into account in the formula. The estimated maximum fictitious $T_c$'s (when $g$ = 1.0) are 29.8 K for Nd$_{1.90}$Ce$_{0.10}$CuO$_4$ at 10 K and 4.2 K for La$_{1.875}$Ba$_{0.125}$CuO$_4$ at 5 K. The estimated $T_c$'s are much higher than their actual $T_c$'s. At least we can say that the extracted $\alpha^2F(\omega)$ may be strong enough to produce the superconductivity if the retarded electron-boson interaction contributes to the paring formation in the materials. We obtained some useful information from the measured data by applying our proposed model.

\section{Discussions and conclusions}

We have shown that we could realize a hump structure in the optical scattering rate within a generalized Allen formalism, which is applicable to a system with non-constant density of states and finite temperatures. We had to introduce a peak followed by a dip structure in the density of states to realize the hump in the scattering rate. The order of the peak and the dip is important because the opposite order result in completely different features in the scattering rate: while the peak-dip order produces a hump structure, the dip-peak order gives a valley. This valley structure in the optical scattering rate can be understood as the normal pseudogap phenomena, which appear near the Fermi energy. However the hump structure, which we can simulate with only peak-dip order, seems to be a completely new physical phenomenon. The physical origin of the dip and/or the peak in the normalized density of states are not understood and clearly justified yet. The dip, which appears after the peak (see in the middle frame of the left column of Fig. \ref{fig5}), is a localized (partial) gap in finite frequency range which may be related to the interband transition observed by recent angle resolved photoemission spectroscopy study of electron doped Sm$_{1.86}$Ce$_{0.14}$CuO$_4$\cite{parkSR:2007}. It may also be closely related to the charge or spin density waves in material systems\cite{dumm:2002,onose:2004,homes:2012} since those systems show hump structures in their optical scattering rates. Since the stripe phase seems to appear in a small doping and temperature region in the hole doped phase diagram, the peak-dip feature in DOS is related to this special phase and is obviously different from the generic pesudogap phase, which is closely related to the dip-peak feature in the DOS.

We also calculated corresponding optical conductivity and reflectance spectra from the simulated optical scattering rate. The resulting conductivity and reflectance spectra are similar to the observed conductivity and reflectance spectra of electron doped cuprates\cite{onose:2004,zimmers:2005,wang:2006} (see in the left column of Fig. \ref{fig4}) and hole doped copper oxides\cite{dumm:2002,homes:2012} (see in the middle column of Fig. \ref{fig4}). We applied this peak-dip model to two measured optical scattering rates with hump features and extracted a reasonable electron-boson spectral function (see in Fig. \ref{fig5}). Many researchers in the high-temperature superconductivity community believe that the retarded electron-boson spectral function may carry the information on the glue for electron-electron cooper pairs, although others suggest that nonretarded interaction also contribute to the paring interaction\cite{Mansart:2013}. This study makes it realistic to extract the electron-boson function out of the optical scattering rate of copper oxides with a hump feature, which was not previously possible. We expect that this work may allow researchers to go one step further in figuring out the superconducting mechanism of cuprates.

\ack

The author acknowledges financial support from the National Research Foundation of Korea (NRFK Grant No. 20100008552 and No. 2012R1A1A2041150). The author thanks J. P. Carbotte for his encouragement for this study, C. C. Homes for sharing his LBCO data for the analysis, and Mr. S. D. Yi at SKKU for upgrading the maximum entropy C$^{++}$ code.

\section*{References}
\bibliographystyle{unsrt}
\bibliography{bib}

\end{document}